\documentclass[pra, 12 pt]{revtex4}

\usepackage[english]{babel}
\usepackage{amsmath}
\usepackage{epsfig}

\newcommand{\tr}{\mathop{\rm tr}\nolimits}

\newcommand{\Pexp}{\mathop{\rm Pexp}\nolimits}

\newcommand{\inst}{\mathop{\rm inst}\nolimits}
\newcommand{\NP}{\mathop{\rm NP}\nolimits}

\newcommand{\eff}{\mathop{\rm eff}\nolimits}

\newcommand{\inputEps}[3]{

\centerline{\epsfxsize=#1 \epsfbox{#3} } \vskip 12pt
{\center\small{#2}} \vskip 18pt}

\begin{document}

\title{Longitudinal electric field: from Maxwell equation to non-locality in time and space}

\author{S.A.Trigger$^{1,2}$}
\address{$^1$Joint\, Institute\, for\, High\, Temperatures, Russian\, Academy\,
of\, Sciences, 13/19, Izhorskaia Str., Moscow\, 125412, Russia;\\
$^2$ Institut f\"ur Physik, Humboldt-Universit\"at zu Berlin,
Newtonstra{\ss}e 15, D-12489 Berlin, Germany;\\
email:satron@mail.ru}

\begin{abstract}
In this paper we use the classical electrodynamics to show that the Lorenz gauge can be incompatible with some particular solutions of the d'Alembert equations for electromagnetic potentials. In its turn, the d'Alembert equations for the electromagnetic potentials is the result of application of the Lorenz gauge to general equations for the potentials. The last ones is the straightforward consequence of Maxwell equations. Since the d'Alembert equations and the electromagnetic potentials are necessary for quantum electrodynamics formulation, one should oblige to satisfy these equations also in classical case.
The solution of d'Alembert equations, which modifies longitudinal electric field is found. The  requirement of this modification follows from the necessity to satisfy the physical condition of impossibility of instantaneous transferring of interaction in space.

PACS number(s): 41.20.Jb, 11.30 Cp, 12.20.-m, 11.10.Ef\\

\end{abstract}

\maketitle

Although Maxwell equations are the experimentally proved background of classical electrodynamics (see, e.g. [1]), the problems connected with the role and observability of the electromagnetic potentials still exist. These problems include, in particular, the various aspects of the Aharonov-Bohm (AB) effect [2], which attracts great attention up to present [3-6]. Most studies, including experimental researches, are devoted to the magnetic AB effect (see, e.g., [7]).

However, in recent years the interest in the electric AB effect has significantly increased (see [8, 9] and references therein), which is due to different interpretations of the experiments [10,11].

The problem of the electric AB effect concerns the basic concepts of the electromagnetic field gauge theory [12]. Since the magnetic field is completely defined by the transverse part of the vector potential, the magnetic AB effect is in fact independent of the gauge invariance of the electrodynamic equations [13]. At the same time, feasibility of the electric analogue of the AB effect creates an apparent contradiction with electrodynamics postulates.

One of the main arguments against the physical reality of the electromagnetic potential is the indeterminacy of its gauge choice. The physical reality of some variable means that its average value (at least for some ranges of the parameters and arguments on which this variable depends) can be single-valued to be reproduced in some experiment.

As is known, the gauge (gradient) invariance is connected with the property of the Maxwell equations for the electromagnetic field to remain unchanged when adding a four-dimensional gradient $\partial \chi/\partial x^\mu$ of an arbitrary scalar function $\chi$ of space--time coordinates. The requirement of relativistic invariance slightly narrows the arbitrariness in the choice of this function (the so-called Lorentz gauge). However, the use of the Lorentz gauge creates significant problems in electromagnetic field quantization, since the electromagnetic potential components become dependent on each other (see, e.g., [14]). But if only the electromagnetic field (i.e., electric and magnetic field strengths) is accepted as a physical reality, then the requirement for relativistic invariance of the potential seems to become excessive. On this basis the Coulomb and other obviously non-covariant gauges are used in many studies.
Recently, in [15] it was considered opportunity to avoid the problem of the Lorenz gauge for operators. The Lorenz gauge has been formulated for averages potentials and on this way the Maxwell equations for averages were justified.

In this paper we use the classical electrodynamics to show that the Lorenz gauge can be incompatible with the particular solutions of the d'Alembert equations for the electromagnetic potentials. In fact, even the Li$\acute{e}$nard -– Wiechert potentials are subordinate to Lorenz gauge only for the case of special choice of the initial conditions [16]. In its turn, the d'Alembert equations for the electromagnetic potentials is the result of the application of Lorenz gauge to the general equations for the potential. These equations are the straightforward consequence of Maxwell equations [16]. Since the d'Alembert equations and the electromagnetic potentials are necessary to formulate the quantum electrodynamics one should oblige to satisfy these equations also in classical case.

This means that the Maxwell equation for the longitudinal electric field behove to modify if the d'Alembert equations are considered as the background of electrodynamics. This requirement also follows to satisfy the physical condition of impossibility of instantaneous transferring of interaction in space.

Starting from the d'Alembert equations
\begin{eqnarray}
\frac{1}{c^2} \frac{\partial^2 {\bf A}}{\partial t^2}- \triangle {\bf A}  = \frac{4\pi {\bf j} }{c}\;, \qquad
\frac{1}{c^2} \frac{\partial^2 {\varphi}}{\partial t^2}- \triangle \varphi  = 4\pi\rho, \label{1}
\end{eqnarray}
we consider the potentials for moving particles in momentum space. The microscopical charge and current densities reads as
\begin{eqnarray}
4\pi\rho (\textbf{r}, t)= 4\pi \sum_a e_a\delta(\textbf{r}-\textbf{r}_a(t))=4\pi\sum_a e_a \int \frac{d^3 k}{(2\pi)^3} \exp[i \textbf{k} (\textbf{r}-\textbf{r}^0_a-\int^t_0 dt'\textbf{v}_a(t')].\nonumber\\
\frac{4\pi {\bf j}(\textbf{r},t) }{c}=\frac{4\pi}{c}\sum_a e_a \textbf{v}_a(t)\delta(\textbf{r}-\textbf{r}^0_a-\int_0^t \textbf{v}_a(t') dt').\qquad
\label{2}
\end{eqnarray}

Equations for the Fourier-components $\varphi_k(t)$ and $A_k(t)$ reads
\begin{eqnarray}
 \frac{\partial^2 {\varphi}_k(t)}{\partial t^2}+ c^2k^2 \varphi_k(t)  = 4\pi c^2\rho_k(t), \label{3}
\end{eqnarray}
\begin{eqnarray}
 \frac{\partial^2 \textbf{A}_k(t)}{\partial t^2}+ c^2k^2 \textbf{A}_k(t) = 4\pi c \textbf{j}_k(t), \label{4}
\end{eqnarray}
where
\begin{eqnarray}
\frac{\partial\rho(\textbf{r},t)}{\partial t}+div \textbf{j}(\textbf{r},t)=0,\; \frac{\partial\rho_k (t)}{\partial t}=-i\textbf{k} {\bf j}_k(t); \nonumber\\
4\pi\rho_k(t)= \frac {1}{2 \pi^2}\sum_a e_a  \exp[-i \textbf{k} (\textbf{r}^0_a+\int^t_0 dt'\textbf{v}_a(t')],\nonumber\\
\frac{4\pi}{c}{\bf j}_k(t)=\frac{ 1 }{2 \pi^2 c}\sum_a e_a \textbf{v}_a(t) \exp[-i \textbf{k} (\textbf{r}^0_a+\int^t_0 dt'\textbf{v}_a(t')].
 \label{5}
\end{eqnarray}

The solutions of these inhomogeneous equations follow from general equation ($w\equiv c k$)
\begin{eqnarray}
 \frac{\partial^2 Y(t)}{\partial t^2}+ w^2 Y (t)  = f(t), \label{6}
\end{eqnarray}
We use below the solution $Y(t)$ ($w=ck$)
\begin{eqnarray}
Y(t)=
\exp (iwt) \int^t dt" \left[\exp(-2iwt") \int^{t"} dt'\exp(iwt')f(t')\right]+C'\exp (iwt)+C"\exp (-iwt). \label{7}
\end{eqnarray}
The first term is one of the particular solutions which we assume physically reasonable.
The terms with the coefficients $C'$, $C"$ provide the possibility of a free longitudinal field existence. Below we neglect these terms which should be specially analysed. 

In Eq. (7) for ${\varphi}_k(t)$ we have take $f(t)$ in the form 
\begin{eqnarray}
f_\varphi(t)=\frac {c^2}{2 \pi^2}\sum_a e_a  \exp[-i \textbf{k} (\textbf{r}^0_a+\int^t_0 dt'\textbf{v}_a(t')].
\label{8}
\end{eqnarray}
For simplicity let us consider the special case of motion with a constant particle accelerations $\textbf{v}(t)=\textbf{v}_b^0+\textbf{a}_b t$. Then $f_\varphi(t)=\frac {c^2}{2 \pi^2}\sum_a e_a  \exp[-i \textbf{k} (\textbf{r}^0_a+\textbf{v}_a^0 t+\textbf{a}_a t^2/2)]$ where index the subscribing index numerates particles.

The general and particular (for the case $\textbf{v}_a(t)=\textbf{v}_a^0+\textbf{a}_a t$ ) forms for ${\varphi}_k(t)$ have the forms
\begin{eqnarray}
\varphi_k(t)=\frac {c^2}{2\pi^2}
\exp (iwt) \int^t dt" \left[\exp(-2iwt") \int^{t"} dt'\exp(iwt')\sum_a e_a  \exp[-i \textbf{k} (\textbf{r}^0_a+\int_0^{t'} d\tau\textbf{v}_a(\tau)\right]
 \label{9}
\end{eqnarray}
and
\begin{eqnarray}
\varphi_{k,a}(t)\frac {c^2}{2\pi^2}\exp (iwt) \int^t dt" \left[\exp(-2iwt") \int^{t"} dt'\sum_a e_a  \exp[-i  (\textbf{k}\textbf{r}^0_a+\textbf{k}\textbf{v}^0_a t'-wt'+\textbf{a}_a t'^2/2)\right],\qquad
 \label{10}
\end{eqnarray}
respectively.
If $\textbf{a}=0$ one can find
\begin{eqnarray}
\varphi_{k,a=0}(t)=-\sum_a
\frac {e_ac^2}{2 \pi^2[(\textbf{k} \textbf{v}_a^0)^2-w^2]}
\exp [-i\textbf{k} \textbf{r}_a^0 -i(\textbf{k v}_a^0)t].\qquad
 \label{11}
\end{eqnarray}

The right side $f(t)$  of Eq.(7) for the potential $\textbf{A}_k(t)$ is the function $f(t)\equiv f_A(t)$
\begin{eqnarray}
f_A(t)=\sum_a
\frac{ e_a c}{2 \pi^2}\textbf{v}_a(t)\exp (-i\textbf{k}\textbf{r}^0_a-i \textbf{k}\int_0^{t} \textbf{v}_a(t") dt") \label{12}
\end{eqnarray}
and the general and particular (for the case $\textbf{v}_a(t)=\textbf{v}_a^0+\textbf{a}_a t$ ) forms of $\textbf{A}_k(t)$
can be written, respectively, as
\begin{eqnarray}
\textbf{A}_k(t)=
\sum_a \frac{ e_a c }{2 \pi^2}\exp (iwt) \times \nonumber\\ \qquad \int^t dt" \left[\exp(-2iwt") \int^{t"} dt'\exp(iwt')\textbf{v}_a(t')\exp (-i\textbf{k}\textbf{r}^0_a-i \textbf{k}\int_0^{t'} \textbf{v}_a(\tau) d\tau)\right]\label{13}
\end{eqnarray}
\begin{eqnarray}
\textbf{A}_{k,a=0}(t)=\sum_a \frac{ e_a c }{2 \pi^2}\exp (iwt) \times \nonumber\\\int^t dt" \left[\exp(-2iwt") \int^{t"} dt'(\textbf{v}_a^0+\textbf{a} t')\exp (-i\textbf{k}\textbf{r}^0_a-i (\textbf{k} \textbf{v}_a^0 t' - wt'+\textbf{k}\textbf{a}_a t'^2/2))\right], \label{14}
\end{eqnarray}
For the case $\textbf{a}=0$
\begin{eqnarray}
\textbf{A}_{k,a=0}(t)=\sum_a
\frac{ e_a c\textbf{v}_a^0}{2 \pi^2}\exp (iwt) \int^t dt" \left[\exp(-2iwt") \int^{t"} dt'\exp [-i\textbf{k}\textbf{r}^0_a+i (w t'-\textbf{k} \textbf{v}_0 t')]\right]=\nonumber\\-\sum_a \frac{ e_a c \textbf{v}_a^0}{2 \pi^2[(\textbf{k}\textbf{v}_a^0)^2-w^2]}
\exp [-i\textbf{k}\textbf{r}^0_a-i(\textbf{k} \textbf{v}_a^0)t].\qquad \label{15}
\end{eqnarray}
Note, that the Lorenz condition for the case $\textbf{a}=0$
\begin{eqnarray}
i \textbf{k}\textbf{A}_{k,a=0}(t)+\frac{1}{c}\frac{\partial \varphi_{k,a=0}(t)}{\partial t}=-\sum_a\frac{i e_a c\textbf{k}\textbf{v}_a^0}{2 \pi^2[(\textbf{k} \textbf{v}_a^0)^2-w^2]}
\exp [-i\textbf{k}\textbf{r}^0_a-i\textbf{k v}_0 t]+\frac{1}{c}\nonumber\\
\sum_a\frac {i(\textbf{k} \textbf{v}_a^0)e_a c^2}{2 \pi^2[(\textbf{k v}_0)^2-w^2]}
\exp [-i\textbf{k}\textbf{r}^0_a-i\textbf{k v}_0 t]=0 \qquad \label{16}
\end{eqnarray}
is fulfilled automatically.

As is easy to see
\begin{eqnarray}
\textbf{E}_k=-i \textbf{k} \varphi_k(t)-\frac{1}{c}\frac{\partial \textbf{A}_k(t)}{\partial t}= \sum_a
\frac {i \textbf{k} e_ac^2}{2 \pi^2[(\textbf{k} \textbf{v}_a^0)^2-w^2]}
\exp [-i\textbf{k} \textbf{r}_a^0 -i(\textbf{k v}_a^0)t]-\nonumber\\i\sum_a \frac{ e_a  \textbf{v}_a^0 (\textbf{k} \textbf{v}_a^0)}{2 \pi^2[(\textbf{k}\textbf{v}_a^0)^2-w^2]}
\exp [-i\textbf{k}\textbf{r}^0_a-i(\textbf{k} \textbf{v}_a^0)t]. \label{17}
\end{eqnarray}
and, taking into account (5) and equality $w^2=k^2c^2$ we arrive at the Maxwell equation
\begin{eqnarray}
i\textbf{k}\textbf{E}_k= - \sum_a
\frac { e_a k^2c^2}{2 \pi^2[(\textbf{k} \textbf{v}_a^0)^2-w^2]}
\exp [-i\textbf{k} \textbf{r}_a^0 -i(\textbf{k v}_a^0)t]+\nonumber\\\sum_a \frac{ e_a  \textbf{k}\textbf{v}_a^0 (\textbf{k} \textbf{v}_a^0)}{2 \pi^2[(\textbf{k}\textbf{v}_a^0)^2-w^2]}
\exp [-i\textbf{k}\textbf{r}^0_a-i(\textbf{k} \textbf{v}_a^0)t]=4\pi \varrho_k(t) . \label{18}
\end{eqnarray}
Therefore, for the case of steady particle motion Maxwell equation for the longitudinal field is fulfilled and the velocity of light is not present in the function $\textbf{k} \textbf{E}_k(t)$.

However, for the case of a non-steady motion the Lorenz condition in general is not valid. The simplest way to check this straightforward is to consider the case of particle motion with a fixed particle accelerations $\textbf{a}_c$, where $c$ is the particle index.
The integrals in (10) and (14) in linear approximation on $\textbf{a}_c$ can be calculated analytically. The respective solution for the potentials shows violation of the Lorenz condition.

Let us determine now the general form of the electric field $\textbf{E}_k(t)$, taking into account that $w=ck$
\begin{eqnarray}
\textbf{E}_k(t)= -i\textbf{k}\varphi_k(t)-\frac{1}{c} \frac{\partial \textbf{A}_k(t)}{\partial t}=\qquad \nonumber\\-i \textbf{k}\frac {c^2}{2\pi^2}
\exp (iwt) \int^t dt" \left[\exp(-2iwt") \int^{t"} dt'\exp(iwt')\sum_a e_a  \exp[-i \textbf{k} (\textbf{r}^0_a+\int_0^{t'} d\tau\textbf{v}_a(\tau)\right]\nonumber\\-
 \frac{\partial}{\partial t}\sum_a \frac{ e_a c }{2 c \pi^2}\exp (iwt)\times \nonumber\\ \int^t dt" \left[\exp(-2iwt") \int^{t"} dt'\exp(iwt')\textbf{v}_a(t')\exp (-i\textbf{k}\textbf{r}^0_a-i \textbf{k}\int_0^{t'} \textbf{v}_a(\tau) d\tau)\right]\qquad
\label{19}
\end{eqnarray}
Let us take into account the discontinuity Eq.(5) to simplify (19)
Then for the longitudinal field we obtain
 \begin{eqnarray}
\textbf{k} \textbf{E}_k(t)= -i k^2\varphi_k(t)-\frac{1}{c} \frac{\partial \textbf{kA}_k(t)}{\partial t}=\nonumber\\-i 4\pi c^2 k^2
\exp (iwt) \int^t dt" \left[\exp(-2iwt") \int^{t"} dt'\exp(iwt')\rho_k(t')\right]+\nonumber\\
4 \pi \frac{\partial}{\partial t}\exp (iwt) \int^t dt" \left[\exp(-2iwt") \int^{t"} dt'\exp(iwt')\frac{\partial \rho_k(t')}{\partial t'}\right]
\label{20}
\end{eqnarray}
If $\rho_k(t')\simeq const$ and accounting that $w=ck$
\begin{eqnarray}
\textbf{k} \textbf{E}_k(t)=-i 4\pi c^2 k^2\rho_k(t)
\exp (iwt) \int^t dt" \left[\exp(-2iwt") \int^{t"} dt'\exp(iwt')\right]=-i4\pi \rho_k(t)
\label{21}
\end{eqnarray}
we arrive at Maxwell equation for the longitudinal field
\begin{eqnarray}
i\textbf{k} \textbf{E}_k(t)=4\pi \rho_k(t)\rightarrow div \textbf{E}(\textbf{r}, t)=4\pi \rho(\textbf{r},t)
\label{22}
\end{eqnarray}

Returning to (20) and taking into account that $ck=w$ we arrive at the expression for longitudinal field
\begin{eqnarray}
\textbf{k} \textbf{E}_k(t)=-i 4\pi w^2
\exp (iwt) \int^t dt" \left[\exp(-2iwt") \int^{t"} dt'\exp(iwt')\rho_k(t')\right]+4\pi \rho_k(t)+\nonumber\\
i 4\pi w \exp (iwt)\{ \int^t dt" \left[\exp(-iwt")\rho_k(t")-iw\exp(-2iwt")\int^{t"} dt'\exp(iwt') \rho_k(t')\right]-\nonumber\\ \exp(-2iwt)\int^{t} dt'\exp(iwt') \rho_k(t')\} \qquad
\label{23}
\end{eqnarray}
or in the equivalent form
\begin{eqnarray}
\textbf{k} \textbf{E}_k(t)=4\pi\rho_k(t)+(1-i) 4\pi w^2
\exp (iwt) \int^t dt" \left[\exp(-2iwt") \int^{t"} dt'\exp(iwt')\rho_k(t')\right]+\nonumber\\
4\pi iw \exp (iwt)\left[ \int^t dt" \exp(-iwt")\rho_k(t")- \exp(-2iwt)\int^{t} dt'\exp(iwt') \rho_k(t')\right].\qquad
\label{24}
\end{eqnarray}
If $\rho_k(t)$ is independent (or slowly dependent) on $t$ we naturally return gain to Maxwell equation (22).

Due to linearity of the relation between the longitudinal field and density, the obtained microscopic equation can be averaged and possesses the same form for the average field and charge density.
The same equation for the average longitudinal field can be reproduced for quantum system of particles and quantum fields on the basis of second quantization approach.

As a result, we arrive at the conclusion that if the d'Alembert equation for microscopical values of the four-vector potential of the electromagnetic field is accepted as the theoretical basis, the Lorentz condition cannot be realized for any particular solution for potentials and arbitrary motion of charges.  The proposed generalization of Maxwell equation for the longitudinal field takes into account the finite speed of the interaction transferring. Only in the limit of slowly changing in time density of charges, the Maxwell local form of the equation for longitudinal electric field is valid.

In contrast to the Maxwell equations for the average longitudinal electric field, the equations for the potentials permit solutions in the form of longitudinal waves (longitudinal and scalar photons), as well as for transversal ones. The detail consideration of this phenomenon as well as possibility of experimental verification of longitudinal quants will be considered in a separate work.

\section*{Acknowledgment}
The author is thankful to V.B. Bobrov and A.I. Ershkovich for the fruitful discussions.
This study was supported by the Russian Science Foundation, project No.14-50-00124.

\end{document}